\documentclass[12pt]{article}
\usepackage{graphicx,float,subfigure,epsfig,color,rotating,latexsym}
\usepackage[latin1]{inputenc}
\def\be{\begin{equation}}
\def\ee{\end{equation}}
\def\bea{\begin{eqnarray}}
\def\eea{\end{eqnarray}}
\def\ba{\begin{array}}
\def\ea{\end{array}}

\def\ln{{\rm ln}}

\def\nin{\noindent}

\begin{document}

\begin{center} {\bf \large{Dilution of zero point energies in the cosmological expansion} }\\

\vspace*{0.8 cm}

Vincenzo Branchina\footnote{vincenzo.branchina@ct.infn.it}\label{one}

\vspace*{0.4 cm}

Dipartimento di Fisica e Astronomia, Universit\`a di Catania \\ and \\ INFN,
Sezione di Catania, Via Santa Sofia 64, I-95123, Catania, Italy

\vspace*{0.8 cm}

Dario Zappal\`a\footnote{dario.zappala@ct.infn.it}\label{three}

\vspace*{0.4 cm}

INFN, Sezione di Catania, Via Santa Sofia 64, I-95123, Catania, Italy

\vspace*{1.2 cm}

{\LARGE Abstract}\\
\end{center}
\nin
The vacuum fluctuations of all quantum fields filling the universe are supposed
to leave enormous energy and pressure contributions which are incompatible with
observations. It has been recently suggested that, when the effective nature of
quantum field theories is properly taken into account, vacuum fluctuations behave
as a relativistic gas rather than as a cosmological constant. Accordingly,
zero-point energies are tremendously diluted by the universe expansion but provide
an extra contribution to radiation energy. Ongoing and future cosmological 
observations could offer the opportunity to
scrutinize this scenario. The presence of such additional
contribution to radiation energy can be tested by using primordial
nucleosynthesis bounds or measured on Cosmic Background Radiation anisotropy.

\vspace*{1.3cm}

Any phenomenological quantum field theory is an effective
field theory (EFT) having a limited energy range of validity.
Quantum electrodynamics as well as the 4-fermion theory of weak
interactions, for instance, are low energy manifestations of a
higher energy theory, namely the Standard Model
(SM), which in turn is the low energy limit of an even higher
energy model. Nowadays, a large amount of experimental and
theoretical work is devoted to the search of the
ultraviolet (UV) completion of the SM.

An EFT is intrinsically defined with a physical UV cut-off
$\Lambda$, the scale which signals the onset of ``new physics''.
Above $\Lambda$ the theory is no longer valid and has to be
replaced by a higher energy one. This results in a hierarchy of
theories each having higher and higher UV
cut-off\,\cite{wilsrep}. This hierarchical structure of elementary 
particle models is belived to survive up
to the Planck scale $M_P$. In the present work, in order to take 
a more general point of a view, we will keep working with a 
generic cut-off $\Lambda$.

This naturally leads to the idea that, whatever theory describes
physics before a specific time $t_\Lambda$, for $t>t_\Lambda$
physics is appropriately described by an EFT with physical cutoff
$\Lambda$. It has been recently shown that a generic consequence
of the effective nature of this EFT (irrespectively of the
specific model) is that the equation of state (EOS) for the vacuum
energy has the unexpected form $p_{vac} = \rho_{vac}/3$, as
for a gas of relativistic particles\,\cite{noi}.

This can be illustrated by considering a single component real
scalar field theory
\be\label{lagra} S[\phi]= \int d^4\,x\,\left
(\frac12\partial_\mu\phi \,
\partial^\mu\phi -\frac12m^2\phi^2 \right )
\ee
and computing the quantum-statistical average (the thermal average
for a statistical equilibrium distribution) of the corresponding
energy-momentum operator,
\be\label{phiemt} T_{\mu\nu} =
\partial_\mu\phi \, \partial_\nu\phi- g_{\mu\nu}{\cal L}
=\partial_\mu\phi \, \partial_\nu\phi - g_{\mu\nu} \left
(\frac12\partial_\sigma\phi \,
\partial^\sigma\phi -\frac12m^2\phi^2 \right )\,.
\ee
In fact, for the vacuum contributions to the averages $<T_{_{0\,0}}>$
and $<T_{ii}>$ (the non-diagonal terms vanish) we find (see \cite{noi}
for details):
\bea\label{diagemt}
&&<T^{vac}_{_{0\,0}}>\, =
\, \frac{1}{V} \sum_{\vec k} \frac{\sqrt{\vec k^2+m^2}}{2}\, \\
&&<T^{vac}_{ii}>\, =  \frac{1}{V} \sum_{\vec
k}\frac{(k^i)^2}{2\,\sqrt{\vec k^2+m^2}}\, , \label{diagemt1} \eea
where $V$ is the quantization volume. By performing in
Eqs.(\ref{diagemt}) and (\ref{diagemt1}) the sum over ${\vec k}$
up to the UV cutoff $\Lambda$, the vacuum contributions
to the energy density $\rho_{vac}= <T^{vac}_{_{0\,0}}>$ and
pressure $p_{vac}= <T^{vac}_{_{i\,i}}>$ (due to rotational
invariance $<T_{_{11}}>=<T_{_{22}}>=<T_{_{33}}>$) turn out to be:

\bea\label{rhov} \rho_{\,vac}    =\frac{1}{16\pi^2}
\left[\Lambda(\Lambda^2+m^2)^{\frac32}-\frac{\Lambda m^2
(\Lambda^2+m^2)^{\frac12}}{2} -\frac{m^4}{4}
\ln\left(\frac{(\Lambda+
(\Lambda^2+m^2)^{\frac12})^2}{m^2}\right)\right]\,, \label{rho} \eea
\bea\label{pv} p_{\,vac}  =\frac{1}{16\pi^2}\left[
\frac{\Lambda^3(\Lambda^2+m^2)^{\frac12}}{3} -\frac{\Lambda m^2
(\Lambda^2+m^2)^{\frac12}}{2} +\frac{m^4}{4}
\ln\left(\frac{(\Lambda+
(\Lambda^2+m^2)^{\frac12})^2}{m^2}\right)\right]\,. \label{press}
\eea

By noting that $\Lambda >> m$, we immediately see that in Eqs.\,(\ref{rhov}) and
(\ref{press}) the dominant terms are the first ones, so we
have: 
\be \label{mainv} 
p_{\,vac} \sim  \frac{\,\,\rho_{\,vac}}{3}\,. 
\ee

It could be objected that the divergences have no physical meaning and
that the definition of a theory has to be completed by some appropriate
renormalization procedure which allows to remove them, so that
$\Lambda$ should be regarded just as a regulator.

As mentioned above, however, from a deeper physical point of
view, it is more satisfactory to consider a quantum field theory 
as an effective theory {\it intrinsically} 
defined with the help of an UV cut-off $\Lambda$ and valid up to this 
scale (see for instance\,\cite{lepage}).
From this perspective, the cut-off $\Lambda$ is physical and the
consequences of its presence in the very definition of the theory have to
be seriously taken into account.

In this respect, it should be recalled that the physical meaning
of the quartic divergence $\Lambda^4$ which originates from
zero point energies is deeply rooted in the underlying harmonic
oscillator structure of a quantum field theory. As pointed out 
in\,\cite{dewitt}, if we cancel out those terms with the help
of a formal procedure such as normal ordering, this property is
automatically lost.

In passing, it is interesting to notice that in recent years it has 
been claimed that dark energy, more specifically zero point energies of 
fundamental fields, can be (and has been) observed in experiments 
which measure the spectral function of the noise current in resistively 
shunted Josephson junctions\,\cite{bema1,bema2}. 
A thorough analysis of the problem, however, has shown that these 
claims were based on a misunderstanding of the physical origin of 
the spectral function\,\cite{jetz1,jetz2,doran,maha,ma1,ma2}. 
In the present work, on the contrary, the footprint of zero point 
energies will be searched for in a different class of experiments
(WMAP, Planck).

Going back to our original problem, we note that the importance 
of the quantum field theoretic
contribution to the energy-momentum tensor that appears in the
Einstein equations,
\be\label{einstein} G_{\mu\nu} - \Lambda_{_{CC}}\, g_{\mu\nu}= 8\,\pi\,
G \,T_{\mu\nu}\, , \ee
was firstly recognized in\,\cite{zeldov} and\,\cite{zeldov2},
where however, in accordance with the idea that the divergences are
unphysical and have to be discarded, the divergent terms were
removed with the help of a renormalization procedure
(Pauli-Villars regulators were used in\,\cite{zeldov2}). Such a
formal approach is thoroughly analyzed and criticized by De
Witt\,\cite{dewitt}. Still, a popular prescription (often used
nowadays) for the automatic (yet formal) cancellation of these
divergences is the dimensional regularization scheme (see, for
instance,\,\cite{birrdav}).

In view of the above considerations, in the present paper we
follow the approach of\,\cite{noi} and adopt the following ``EFT
point of view'': we assume that at a given time $t_\Lambda$ physics
is described by an EFT with cut-off $\Lambda$. 

Scope of this work is to  explore some of the 
phenomenological consequences of this assumption. In particular, 
we show that the resulting scenario does not conflict with well 
known results of standard cosmology and that in the very near future 
it will be possible to test it against experimental data.

Actually, when the renormalized (as opposed to effective) theory
is considered, i.e. when the divergent terms are discarded with
the help of a specific renormalization scheme (dimensional
regularization, Pauli-Villars regulators, ...), the coefficient
$w$ in the EOS \, $p_{vac}=w\,\rho_{vac}$\, 
turns out to be $w=-1$\,\,\cite{zeldov2,birrdav} (see also the above
Eqs.\,(\ref{rhov}) and (\ref{press}) with the quartic and
quadratic divergent terms subtracted). We then conclude that
$\rho_{vac}$ does not evolve with time and can be interpreted as a
contribution to the cosmological constant. This is the standard
view.

Within our effective field theory approach, on the contrary, we
keep the large but finite terms proportional to $\Lambda^4$ and
$\Lambda^2$ and therefore we obtain a different EOS,
Eq.(\ref{mainv}), with $w \sim
1/3$, which is the same as for the relativistic matter case. The
immediate implication of this is that the zero point energy
density of a quantum field red-shifts with time.

This result suggests the possibility of considering the following
scenario, the theoretical aspects of which have been studied
in\,\cite{noi}.

First of all, we assume that at the time $t \sim t_\Lambda$ and
energy scale $E\sim \Lambda$ physics is entirely described by a
quantum field, a scalar field for example, and that the 
lower energy theories were born during the cosmic time evolution.
This assumption appears natural in view of our ideas on the
effective nature of particle physics theories and fits our current
views on the cosmological evolution. Accordingly, the zero point 
energy density associated to the quantization of this field at 
$t=t_\Lambda$ is 
\be \label{zpen} \rho_{vac}(t_\Lambda)=\frac{\Lambda^4}{16\pi^2}
\label{eq:zero} \ee
where again we have taken $\Lambda >> m$. 

Second we note that, from this perspective, the lower energy fields,
i.e. lower energy degrees of freedom (dof), are nothing but a convenient
way to parametrise the theory at a lower scale. Therefore, when
computing the vacuum contribution to the energy density, one should
not include the zero point energies of the effective low energy theories
as this would result in a multiple counting of dof. The only contribution
from zero point energies to the energy density of the universe comes
from the dof of the original theory. 

According to this scenario, due to Eq.\,(\ref{mainv}), the zero-point 
energy contribution to the energy density of the universe, which at 
$t_\Lambda$ is\,\, $\rho_{vac}(t_\Lambda)\,\sim\,\Lambda^4$ 
(see Eq.\,(\ref{eq:zero})), is tremendously diluted by the cosmic evolution.
In other words, according to our picture, the zero point energies of an effective 
field theory (with any mass term neglected) are expressed by the parameter $\Lambda$,
which corresponds to the upper energy scale cut-off of the effective theory itself.
Then, due to their EOS, Eq.\,(\ref{mainv}), which is equal to the EOS for radiation, 
the zero point energies undergo the same renowned rescaling 
of radiation energy density with cosmic time.

Clearly, this model does not explain the
origin of the measured dark energy (which has the experimentally
estimated value $w \sim -1$). However it has the virtue of proposing a
natural solution for the theoretical conundrum of the 120 order of
magnitude excess of energy density coming from the vacuum
fluctuations of quantum fields.

At the same time, this red-shifted zero-point energy provides
an additional contribution to the radiation energy density (photons,
neutrinos), as can be expected from the EOS, Eq.\,(\ref{mainv}).

Ongoing (as well as future) cosmological observations (will) offer the 
opportunity to test this scenario against experiments. As we shall 
see, an important result of our present analysis is that such an 
excess of radiation-like energy density does not screw up well 
tested theoretical predictions of standard cosmology and, at the 
same time, can be compared with present and forthcoming experimental 
results.

In order to allow for such a comparison, let us consider the usual 
parametrization for the total amount of radiation energy density that 
we are considering at a generic time $t$ after neutrino
decoupling\,\cite{shva,schra},
\begin{equation} \rho_{rad}(t) = \rho_\gamma(t) + \rho_\nu(t) + \rho_x(t) =
\left(1 + \frac78 \left(\frac{4}{11}\right)^{4/3}
N_{eff}\right)\rho_{\gamma}(t)\,\,, \label{neff}
\end{equation}
with $N_{eff}$ given by $N_{eff} = 3 + \delta n_x$, where $3$ is
the standard number of neutrinos and $\delta n_x$ accounts for
possible extra dof, so called neutrino
equivalent dof. Hence, in order to calculate the contribution of
zero-point energy contribution  to $\delta n_x$, we have to evolve
$\rho_{vac}$ from $t_\Lambda$ down to $t$.

Let us remind that, since $\rho_{vac}$ is a radiation-like term, one
can use the following equation to extrapolate the values to the
present time $t_0$
\be \frac{\rho_{vac}(t)}{\rho_{\gamma}(t)} =
\frac{\rho_{vac}(t_0)}{\rho_{\gamma}(t_0)} \label{eq:l1} \ee
Furthermore, as shown in \cite{noi}, $\rho_{vac}$ at present time
is given  by
\be \rho_{vac}(t_0) = \rho_{vac}(t_\Lambda)
\left(\frac{t_\Lambda}{t_0}\right)^2 \, \frac{1}{1+z_{eq}}
\label{eq:l2}\ee
By using Eqs.\,(\ref{eq:zero}), (\ref{eq:l1}) and (\ref{eq:l2}),
we can write
\be \frac{\rho_{vac}(t)}{\rho_{\gamma}(t)} =
\frac{\rho_{vac}(t_\Lambda)}{\rho_{\gamma}(t_0)}
\left(\frac{t_\Lambda}{t_0}\right)^2
 \, \frac{1}{1+z_{eq}} = \frac{135}{64 \, \pi^4} \,
 \left(\frac{M_P\, H_0}{T_0^2}\right)^2\,\frac{1}{1+z_{eq}} 
\,\left(\frac{\Lambda^2 \, t_\Lambda}{M_P}\right)^2
\label{eq:3}\, ,  \ee
where $z_{eq}$ is the value of the redshift at the time of 
matter-radiation equality, while $H_0$ and $T_0$
are the Hubble constant and the temperature of the 
relic photons at present time $t_0$ respectively. 
Note that all the terms on the r.h.s. of the previous expression are 
known except the product $\Lambda^2 \, t_\Lambda$, which
is the free parameter of our model. By replacing Eq.\,(\ref{eq:3}) 
in Eq.\,(\ref{neff}) and by taking 
$M_P=1.22\, 10^{19} \, GeV$, $T_0=2.725 \, K$, ${H_0}^{-1}=1.27\, 10^{26} \, m$, 
\cite{pdg}, and $z_{eq}=3253$, \cite{zwmap},  we get :
\be\label{bb} 
\delta n_x =  \frac87 \left(\frac{11}{4}\right)^{4/3}\,
\frac{135}{64 \, \pi^4} \,
 \left(\frac{M_P\, H_0}{T_0^2}\right)^2\,\frac{1}{1+z_{eq}} \,\left(\frac{\Lambda^2 \,
 t_\Lambda}{M_P}\right)^2 \sim 3.46 \left(\frac{\Lambda^2 \,
 t_\Lambda}{M_P}\right)^2\,.
\ee

We turn now our attention to the existing experimental constraints to our 
free parameter $\Lambda^2 \, t_\Lambda$. To this end, we refer to 
a recent analysis\,\cite{rino}, where it was shown that 
\be\label{bbb}
\delta n_x=0.18^{+0.44}_{-0.41}
\ee 
at 95 $\%$ C.L.
With the help of Eqs.\,(\ref{bb}) and (\ref{bbb}), we immediately see 
that, within two standard deviations (or by taking the central value), 
we get the upper bound: 
\be\label{last}
\frac{\Lambda^2 \, t_\Lambda}{M_P} \leq 0.42 \,\, (0.23). 
\ee

Eq.\,(\ref{last}) gives an experimental bound for the product  of 
the two parameters $\Lambda^2$ and $t_\Lambda$, 
which are the essential ingredients of Eq.\,(\ref{zpen}). If the 
theory is quantized at $t_\Lambda=t_P={M_P}^{-1}$,
one gets $\Lambda\leq \,0.65\,M_P\,\,\, (0.48\,M_P )$. Remarkably, 
these results agree with the natural choice of taking $\Lambda$ 
at $t_P$ of the same order of magnitude of $M_P$. 

In this respect, it is worth to notice that the choice $\Lambda=M_P$ 
at $t_\Lambda=t_P$, which we used in\,\cite{noi}, falls beyond the 
two standard deviation limit in Eq.\,(\ref{last}). But it must also 
be remarked that this too stringent limit stems from the simplest
possible scenario considered here, namely a single weakly interacting 
scalar field. For more involved models, the numerical 
bound in Eq.\,(\ref{last}) can be slightly modified.

Clearly, Eq.\,(\ref{last}) leaves open  the possibility of quantizing 
the EFT at later times, $t_\Lambda> t_P$, provided that the UV 
cut-off $\Lambda$ is appropriately rescaled in order to fulfill  
the bound. We observe that, according 
to Eq.\,(\ref{last}), $\Lambda$ has to scale with $t_\Lambda$ exactly 
in the same way as the temperature of a radiation dominated universe does with 
cosmic time. This indicates that, at times of the order of $t_P$,
the temperature of the universe is directly related to the energy scale 
cut-off of the quantum field theories describing the universe itself, as 
should be expected on physical grounds.

Interestingly, the extra amount of radiation predicted in our model 
could be appreciable from forthcoming experiments on Cosmic Background
Radiation (CBR) anisotropy, like those of the Planck satellite. 
Their data could be used to explore the possibility of a direct 
measurement of the primordial zero-point energy density of the EFT 
which describes physics at the Planck scale.

Our arguments, although only suggestive, clearly indicate that there is
space for a non standard scenario which could explain the absence
of the enormous amount of energy density predicted by a
straightforward but simplistic field quantization. According to our
scenario, the zero-point energy contribution to the energy density of
the universe is almost entirely washed out by the cosmological
evolution. After the cosmic dilution has taken place, however, there is
an extremely tiny left over which could be experimentally detected 
in the near future as an additional contribution to the radiation energy 
density content of our universe.

Actually, {\it via} their sensitivity to the universe radiation
content at the time of last scattering, new CBR data should provide 
constraints against which our model predicting a washing 
out of zero point energies could be tested. These data might signal 
whether the zero-point energies should be searched  in a previously 
unsuspected sector (radiation) of the universe energy budget.

Before ending this letter, we would like to add few more comments. 
First of all, we would like to stress again that our momentum 
cutoff is not Lorentz invariant (in the more general case it would 
not be diffeomorphism invariant). Lorentz symmetry violating 
effects and related extensions of the Standard Model have been widely 
studied in the past years, see e.g.\,\cite{kost1} 
and, in particular,  it has been shown that these effects can 
be described within the framework of effective field theories\,\cite{kost2}.
In the effective field theory under consideration, the Lorentz violation 
has measurable consequences only as far as the the vacuum energy density 
is concerned and therefore it can be entirely  associated to a property 
of the vacuum state.
From the phenomenological point of view, one should expect to observe
these Lorentz symmetry violating effects in the experimental determination
of the radiation energy density, as an excess with respect to 
standard expectations. However, as we have shown above, these effects 
are very tiny and their possible detection would be the fingerprint of 
the preferred cosmology frame. 

Moreover, it is worth to notice that, while our analysis concerns the 
fate of the zero point energies and 
points towards the cancellation of the latter due to 
Eq.\,(\ref{mainv}), other authors have considered another 
important facet of the cosmological constant problem, namely the 
possibility of a running with the energy scale of the cosmological 
constant term $\Lambda_{CC}$ which appears in the Einstein-Hilbert
quantum gravity action\,\cite{reuter,sola,ward}. 

The latter approach shows some common features with the mechanism described here.
In fact, in both cases it is considered (in the framework 
of the Wilsonian effective theories) the running of some parameters 
entering the Einstein  equations, Eq.\,(\ref{einstein}), 
with an  energy scale that  can be  eventually re-expressed 
in terms of the cosmic time. However, unlike these works
which are focused on the running of the cosmological and Newton constants, 
we are only concerned with a single contribution to the energy-momentum tensor,
namely the zero point energies contribution of an effective field theory.
This  has a different EOS (and therefore a different time evolution) with 
respect to the cosmological constant, with the consequence that the former 
decreases with time and is very tiny (and can be detected only under favourable 
conditions as explained above), whereas
the latter, as confirmed by observations, has a much larger measured value
and a different EOS, namely $w\simeq -1$.

Nevertheless, the running considered by these authors, 
could provide (and we believe it does) the dynamical 
mechanism (suppression by quantum fluctuations) which eventually 
should give the measured value of the CC. In this respect, we think
that our analysis together with the work these authors should naturally
merge in a coherent picture which could provide a scenario 
for the solution of the CC problem in its different aspects.

\vspace{15 pt}

\noindent 
{\large {\bf Aknowledgements}}
\vspace{15 pt}

\noindent 
We would like to thank G.Mangano and G.Miele for several useful discussions.

\end{document}